\documentclass{aa}  
\usepackage{graphicx,natbib}
\bibpunct{(}{)}{;}{a}{}{,}
\usepackage{txfonts}

\begin{document}
 
   \title{Effect of motions in prominences on the helium resonance lines in the extreme ultraviolet}
   \titlerunning{Effect of motions in prominences on helium resonance lines}

   \author{N. Labrosse\inst{1}
          \and P. Gouttebroze\inst{2}
	  \and J.-C. Vial\inst{2}
          }

   \offprints{N. Labrosse}

   \institute{Institute of Mathematical and Physical Sciences, University of Wales Aberystwyth, Ceredigion SY23 3BZ, UK\\
              \email{nll@aber.ac.uk}
         \and
             Institut d'Astrophysique Spatiale, CNRS--Universit\'e Paris Sud, 91405 Orsay cedex, France\\
             \email{pierre.gouttebroze@ias.u-psud.fr,jean-claude.vial@ias.u-psud.fr}
             }

   \date{Received 8 June 2006 / Accepted 4 December 2006}

  \abstract
   {Extreme ultraviolet resonance lines of neutral and ionised helium observed in prominences are difficult to interpret as the prominence plasma is optically thick at these wavelengths. If mass motions are taking place, as is the case in active and eruptive prominences, the diagnostic is even more complex.}
  {We aim at {studying the effect of radial motions on the spectrum emitted by moving prominences in the helium resonance lines and at
     facilitating the interpretation of observations}, in order to improve our understanding of these dynamic structures.}
   {We develop our non-local thermodynamic equilibrium radiative transfer code formerly used for the study of quiescent prominences. The new numerical code is now able to solve the statistical equilibrium and radiative transfer equations in the non-static case by using velocity-dependent boundary conditions for the solution of the radiative transfer problem. This first study investigates the effects of different physical conditions (temperature, pressure, geometrical thickness) on the emergent helium radiation.}
  {The motion of the prominence plasma induces a Doppler dimming effect on the resonance lines of \ion{He}{i} and \ion{He}{ii}. The velocity effects are particularly important for the \ion{He}{ii} $\lambda$\,304~\AA\ line as it is mostly formed by resonant diffusion of incident radiation under prominence conditions. The \ion{He}{i} resonance lines at 584 and 537~\AA\ also show some sensitivity to the motion of the plasma, all the more when thermal emission is not too important in these lines. We also show that it is necessary to consider partial redistribution in frequency for the scattering of the incident radiation.}
   {This set of helium lines offers strong diagnostic possibilities that can be exploited with the SOHO spectrometers and with the EIS spectrometer on board the Hinode satellite. The addition of other helium lines and of lines from other elements (in particular hydrogen) in the diagnostics will further enhance the strength of the method.}


   \maketitle

\defcitealias{lg01}{LG01}

\section{Introduction}

It is still not clear how solar prominences reach a state of equilibrium. However it is known that during the formation stage, significant plasma motions are taking place. Similarly, mass motions are important during the disappearance of prominences. Between these two stages of the life of prominences, one may also observe periods of activity with internal plasma motions \citep[see for example][]{ofmanetal98,ktp03}.
In order to make an accurate assessment of the energy budget in solar prominences, detailed observations of a number of lines formed at different temperatures are crucial. 
{Among them, the helium lines are important, as they are strong and complement the hydrogen lines to reveal the physical conditions of the prominence plasma in different regions of the structure.}
However, extreme ultraviolet resonance lines of neutral and ionised helium are difficult to interpret as the prominence plasma is optically thick at these wavelengths. If mass motions are taking place, the diagnostic is even more complex.
{Furthermore , h}elium is the most abundant element after hydrogen, and therefore significantly contributes to the energy budget, {for instance by absorbing coronal radiation at wavelength below the \ion{He}{i} ionisation threshold ($\lambda < 504$~\AA) -- see, e.g., \citet{vaetal99,ah05}. It does not, however, significantly contribute to the radiation losses \citep{kp91}.
In all cases, it appears that in order to build realistic prominence models, detailed non-LTE radiative transfer modelling is necessary.}

We aim at facilitating the interpretation of observations in the helium resonance lines in moving prominences, in order to improve our understanding of these dynamic structures.
{This paper is thus devoted to the study of the effect of radial motions in prominences on the helium resonance lines.}
We present the first computations of the helium line profiles emitted by a moving prominence. The prominence is modelled as a plane-parallel slab standing vertically above the solar surface and moving upward as a solid body. The helium spectrum is computed with a non local thermodynamic equilibrium radiative transfer code. The modelling aspects are presented in \S~\ref{modelling}. 
The effect of Doppler dimming is investigated in \S~\ref{results} for the EUV resonance lines of \ion{He}{i} at 584~\AA\ and 537~\AA\ and \ion{He}{ii} at 304~\AA. 
We focus on the line profile properties and the resulting integrated intensities {normalised to their respective values when the prominence is at rest}. We also study the effect of frequency redistribution in the line formation mechanisms. We discuss the results in \S~\ref{discuss}.

\section{Modelling of an eruptive prominence}\label{modelling}

\begin{figure}
  \resizebox{\hsize}{!}{\includegraphics{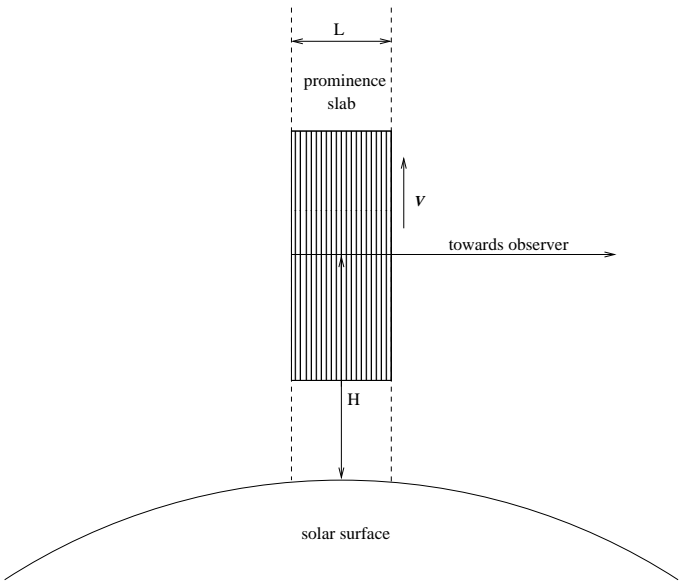}}
  \caption{{Schematic representation of the free-standing vertical prominence slab. The slab has a geometrical extension $L$ along the line-of-sight to the observer, at the altitude $H$ above the limb. The prominence plasma is moving radially outwards at the velocity $\vec{V}$.}}
  \label{fig:promslabv}
\end{figure}

We follow the approach of \cite{gvg97b,gvg97a} who computed the hydrogen spectrum emitted by an eruptive prominence. The prominence is modelled as a {one dimensional plane-parallel slab standing vertically above the solar limb {(Fig.~\ref{fig:promslabv})}. Its geometrical thickness along the line of sight and its altitude above the limb are free parameters. The other free parameters, which define the prominence atmosphere, are the electron temperature, the gas pressure, and the microturbulent velocity. The incident radiation field is identical on both sides of the prominence. Therefore the model is symmetrical and we need to solve all the equations in one half of the slab only.}
For this first study of the helium spectrum emitted by eruptive prominences, we only consider isothermal and isobaric prominence models similar to those presented in \citet[][{hereafter \citetalias{lg01}}]{lg01}, although our code allows the inclusion of a transition region between the cold prominence core and the hot surrounding corona \citep{protuspm10,lg04}.

\begin{figure}
  \resizebox{\hsize}{!}{\includegraphics{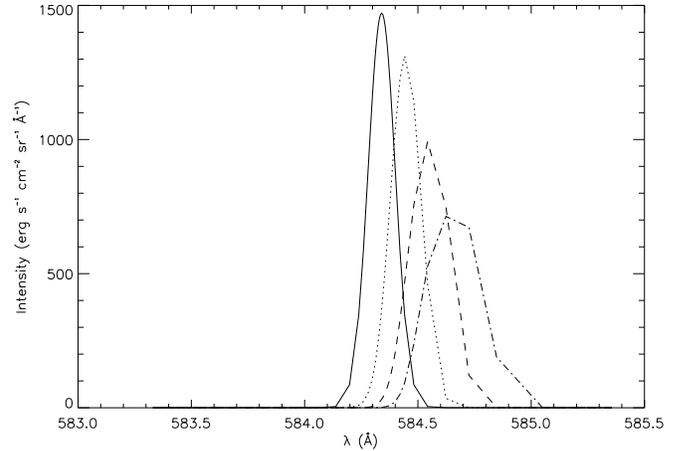}}
  \caption{Angle-averaged incident profiles {in the prominence frame} in the \ion{He}{i} 584 line for a prominence located at 50\,000~km above the limb, and for velocities ranging from 0 (solid line) to 240~km~s$^{-1}$ (step=80~km~s$^{-1}$). The incident profile is shifted towards the red when the velocity increases.}
  \label{incrad}
\end{figure}

\begin{figure*}
  \centering
  \includegraphics[width=15cm]{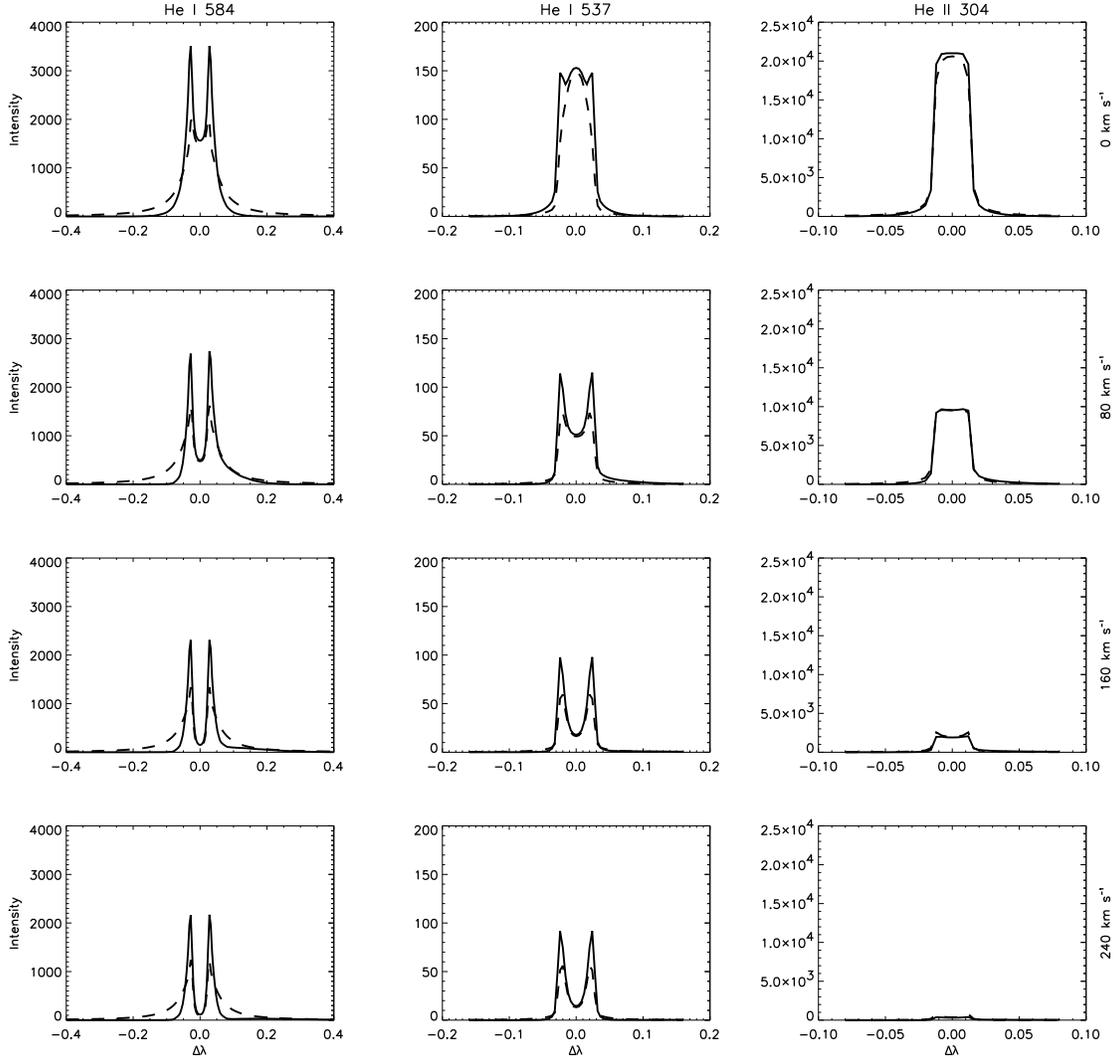}
  \caption{Differences between CRD ({dashed} line) and PRD (solid line) for the profiles of the three helium lines at different velocities: 0, 80, 160, and 240 km\,s$^{-1}$ from top to bottom. Abscissa is $\Delta\lambda$ (distance to line centre) in~\AA\ and vertical axis is specific intensity in erg\,s$^{-1}$\,cm$^{-2}$\,sr$^{-1}$\,\AA$^{-1}$. {Model parameters are altitude $H=50\,000$~km, width $L=650$~km, temperature $T=6500$~K, and pressure $p=0.1$~dyn cm$^{-2}$.}}
  \label{profcrdprd}
\end{figure*}

\begin{figure*}
  \centering
  \includegraphics[width=15cm]{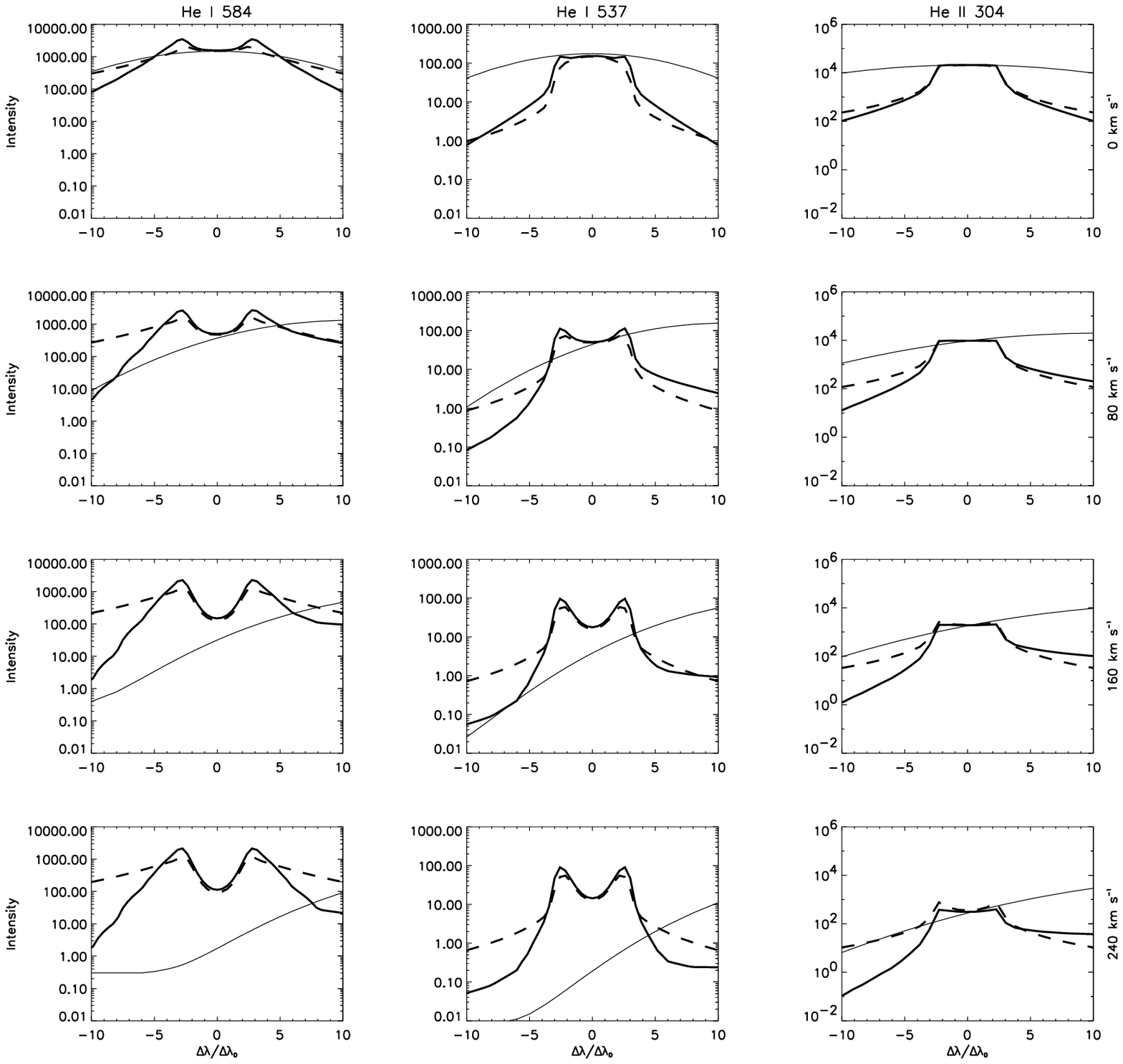}
  \caption{{Same as Fig.~\ref{profcrdprd}, with the abscissa now in Doppler units and limited to 10 Doppler widths around line centre, and the intensities on the vertical axis now shown on a log scale. The thin solid line shows the incident line profile, gradually shifting towards the red as the prominence is moving radially upwards.}}
  \label{profincrad}
\end{figure*}

When the prominence is moving radially outwards, the incident radiation coming from the solar disk and illuminating the structure is shifted to lower frequencies. Therefore we must take the motion of the material into account in the radiative transfer calculations.

The numerical procedure has been described elsewhere \citep[][and references therein]{lg01,lg04} and we sum it up as follows:
\begin{enumerate}
  \item Equations for pressure equilibrium, the ionisation and statistical equilibria of the hydrogen atom are solved, together with radiative transfer for the hydrogen lines and continuum.
  \item Then, statistical equilibrium and radiative transfer equations are solved for other elements. Here we focus on \ion{He}{i} and \ion{He}{ii}. {The helium abundance is fixed at $0.10$.
    The atomic model for He is the same as in LG01. It consists in 29 atomic levels up to $n=5$ for \ion{He}{i}, 4 atomic levels up to $n=4$ for \ion{He}{ii}, and one continuum level for \ion{He}{iii}. Energy levels and statistical weights are given by \cite{wiese66}, while the photoionisation cross-sections come from TOPBASE \citep{topbase_he}. For \ion{He}{i}, collisional ionisation and spontaneous emission coefficients are those of \cite{benj}; effective collision strengths are given by \cite{benj} and \cite{bk72}; and Stark broadening coefficients are taken from \cite{dsb84} and \cite{griem74}. For \ion{He}{ii}, we use the effective collision strengths of \cite{aggarwaletal91,aggarwaletal92}; collisional ionisation coefficients from \cite{ms68}; and the spontaneous emission coefficients are those of \cite{allen73}.}
\end{enumerate}

The velocity-dependent incident radiation illuminating the prominence slab is represented by the mean intensity 
\begin{equation}
  J_0(\nu) = \frac{1}{4\pi} \oint{I_0 \left(\nu+\frac{\nu_0}{c} \vec{V} \cdot \vec{n^\prime}, \vec{n^\prime} \right) \mathrm{d\vec{n^\prime}}} \ ,
\end{equation}
where $I_0(\nu,\vec{n})$ is the specific intensity of the incident radiation, $\vec{n^\prime}$ is the direction of the incident photon, and $\vec{V}$ is the prominence velocity. $J_0(\nu)$ is calculated at a given height, taking into account the center-to-limb variations (if any) of the incident radiation.
In fact we neglect the center-to-limb variations of the helium EUV resonance lines studied here. {In the remaining of this paper we will refer to the angle-averaged, height-diluted incident profile simply as the \textit{incident profile}, unless otherwise stated.}

{The incident radiation in the different lines and continua of hydrogen and helium is taken into account. For helium, we use gaussian incident line profiles for \ion{He}{i} $\lambda$\,584~\AA\ ($I(584)=685$~erg s$^{-1}$ cm$^{-2}$ sr$^{-1}$, FWHM=0.140~\AA), \ion{He}{i} $\lambda$\,537~\AA\ ($I(537)=75$~erg s$^{-1}$ cm$^{-2}$ sr$^{-1}$, FWHM=0.129~\AA), \ion{He}{ii} $\lambda$\,304~\AA\ ($I(304)=7000$~erg s$^{-1}$ cm$^{-2}$ sr$^{-1}$, FWHM=0.100~\AA), and \ion{He}{ii} $\lambda$\,256~\AA\ ($I(256)=167$~erg s$^{-1}$ cm$^{-2}$ sr$^{-1}$, FWHM=0.084~\AA). The incident radiation in the other helium lines is taken as in \citet{hmp}, except for \ion{He}{i} $\lambda$\,6678~\AA\ where the incident intensity is the same as in \citet{sw97}. The incident radiation in the different continua of \ion{He}{i} and \ion{He}{ii} is defined by a brightness temperature table as in \citet{hmp}, but we use EUV fluxes given by \citet{hch74} for wavelengths below 300~\AA.}

{Figure~\ref{incrad}} shows how the incident profile in the first \ion{He}{i} resonance line at 584.3~\AA\ is modified by the outward motion of the medium. The Doppler effect induces a shift of the profile, and the variation of the Doppler shift with direction induces a distortion of the incident profile. 
This effect has already been illustrated in several studies \citep[see for instance][]{hr87,gvg97b}.
As the outward velocity increases, the central peak of the incident profile is less intense, it is moved towards the red, and the line width becomes larger.
{The same effect is also found for all other lines where the incident radiation is frequency-dependent, including the two other lines under study (\ion{He}{i} 537 and \ion{He}{ii} 304). Where no detailed incident profile is included in our code, the incident radiation is taken as constant across the line profile and therefore no Doppler effect arises.}

\section{Results}\label{results}

\subsection{Frequency redistribution}

It is well known that in solar prominences, the redistribution in frequency during the scattering of the incident radiation in resonance lines is best described by a combination of complete frequency redistribution (CRD) and coherent scattering in the atom's rest frame \citep{hgv87}. This leads us to use the standard partial redistribution (PRD) approximation to treat the scattering of the incident radiation. We know from \cite{gvg97b} that for hydrogen emission {in moving prominences,} PRD leads to emerging line profiles which are different than in CRD. Is it also the case for helium?

{We compare two cases. In the CRD case, all hydrogen and helium lines are treated in CRD, while in the PRD case, PRD is used for the hydrogen Lyman lines (Ly$\mathrm{\alpha}$, Ly$\mathrm{\beta}$, and Ly$\mathrm{\gamma}$) and the \ion{He}{i} $\lambda\lambda$\,584 and 537 \AA\ and \ion{He}{ii} $\lambda$\,304 \AA\ resonance lines.
Our angle-averaged
frequency redistribution function is a linear combination of the redistribution functions $R_{\mathrm{IIA}}$ and
$R_{\mathrm{IIIA}}$ introduced by \cite{hummer62}. With the branching ratio $\gamma=\Gamma_r/(\Gamma_r+\Gamma_c)$, where
$\Gamma_r$ and $\Gamma_c$ are the radiative and collisional damping constants
respectively, our PRD redistribution function is \citep{osc72}: 
\begin{equation}
  R(\nu,\nu^\prime) = \gamma R_{\mathrm{IIA}}(\nu,\nu^\prime) + (1-\gamma) R_{\mathrm{IIIA}}(\nu,\nu^\prime) \ .
  \label{defredi}
\end{equation}
$R_{\mathrm{IIIA}}(\nu,\nu^\prime)$ is taken to be equal to the complete redistribution function given by the product $\varphi_\nu \varphi_{\nu^\prime}$ as in \cite{mihalas}, where $\varphi_\nu$ is the normalised absorption profile of the line, and $\nu$ and $\nu^\prime$ are the frequencies of the absorbed and re-emitted photons.
Since the Lyman~$\mathrm{\alpha}$ line has extended wings, we use a frequency-dependent
collisional damping coefficient \citep{yeletal81}. We impose $\gamma=0$ in
eq.~(\ref{defredi}) for lines treated
with the {standard} CRD approximation.}

{To investigate the effects of frequency redistribution we consider one prominence model with the following parameters: temperature $T=6500$~K, pressure $p=0.1$~dyn\,cm$^{-2}$, width $L=650$~km, altitude $H=50\,000$~km, and no microturbulent velocity. This is similar to the models used by \cite{hr87} and \cite{gvg97b}. 
We present the emergent profiles of the \ion{He}{i} $\lambda\lambda$\,584 and 537 \AA\ and \ion{He}{ii} $\lambda$\,304 \AA\ resonance lines computed in CRD (dashed lines) and in PRD (solid lines) at different velocities (0, 80, 160, and 240 km\,s$^{-1}$) in figures \ref{profcrdprd} and \ref{profincrad}. Fig.~\ref{profincrad} is a close-up of Fig.~\ref{profcrdprd} for $\lambda_0 - 10\lambda_D \leq \lambda \leq \lambda_0 + 10\lambda_D$, with the abscissa given in units of the Doppler width $\lambda_D$ of each line instead of angstr\"oms, and the intensities shown on a logarithmic scale.}

Fig.~\ref{profcrdprd} shows that the \ion{He}{i} resonance line profiles differ between the CRD and PRD cases at all velocities.
First of all, the profiles computed with the CRD approximation lead to symmetrical line profiles at all speeds. On the contrary, PRD leads to quasi-coherent scattering in the line wings. It can then reproduce asymmetries that may be present in the incident line profile. When the prominence is at rest, our incident line in these calculations has a symmetrical shape and therefore, the emergent line profile is symmetrical as well. However, when the prominence is moving, the incident radiation is stronger in the red than in the blue parts of the line (see for instance the shape of the incident profile at 584 \AA\ at different speeds in Fig.~\ref{incrad}). Thus the quasi-coherent scattering of the line-wing photons leads to higher intensities in the red wings than in the blue wings of the emitted lines. 
One can also note that the line self-reversal is more pronounced in PRD than in CRD, the line peaks having a higher intensity in the former than in the latter case while the intensity at line centre is the same. The peak intensities of the \ion{He}{i} resonance lines are found between 2 and 3 Doppler widths from line centre. As far as the \ion{He}{ii} line is concerned, there is no obvious difference in the line profiles computed in CRD and in PRD when shown on a linear scale. Here again there are asymmetries in the wings of the PRD profiles that do not exist in CRD and which are best seen on a log scale.

Finally, one can see that asymmetries are significant in the profile of the 584~\AA\ line at low speeds (say less than 150 km\,s$^{-1}$) in the PRD case. In fact, the intensity in the far wings increases with speed in the red, and decreases with speed in the blue, thus enhancing the asymmetry of the line profile as the velocity gets higher. The line asymmetries are more obvious for this line because the wings contribute for the largest part to the emitted radiation, as  discussed {below.}
The asymmetry in the {PRD} profile of the \ion{He}{ii} line also increases with the velocity of the prominence material, although the low intensity in the line wings makes this fact less visible.

In addition to the emergent line profiles within 10 Doppler widths of the central wavelength, we show in Fig.~\ref{profincrad} part of the incident line profile (thin solid line). The comparison of the incident and emergent line profiles on Fig.~\ref{profincrad} unveils the formation mechanisms for the resonance lines of \ion{He}{i} and \ion{He}{ii}. The latter line is the simplest to understand. The fact that the emergent intensity at line centre is exactly the incident intensity times the dilution factor reflects what we already know, namely that this line is solely formed by the scattering of the incident radiation. Note that this is true at all velocities considered here. 
This makes the \ion{He}{ii} 304 line an excellent proxy for the diagnostic of radial speeds in eruptive prominences, and this is confirmed in the rest of this paper.

When the prominence is at rest, the emergent intensity of the first \ion{He}{i} resonance line at 584 \AA\ at line centre is the same as the incident intensity (with the dilution factor taken into account), again an indication that the scattering of the incident radiation plays a major role in the formation of the line. However, when the prominence velocity is increasing the emergent intensity at line centre is higher than the incident intensity, which implies that there are other processes at work. These other processes mainly consist of collisional excitations and coupling with other transitions (principally between the singlet states of the \ion{He}{i} system).

\begin{figure}
  \resizebox{\hsize}{!}{\includegraphics{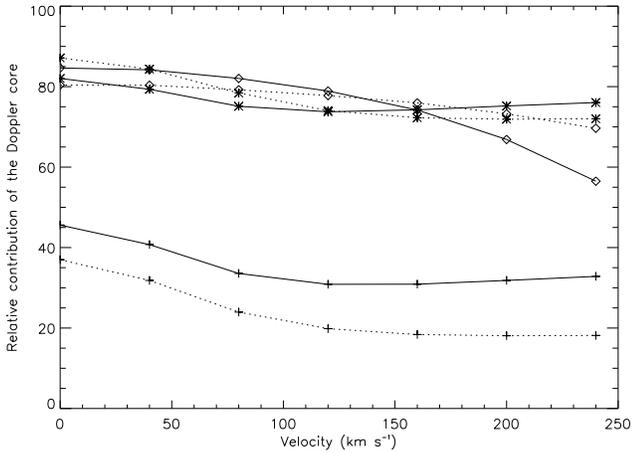}}
  \caption{Comparison of the relative contribution of the Doppler core in the emitted line integrated intensities between CRD (dotted line) and PRD (solid line) calculations for \ion{He}{i} $\lambda\lambda$\,584 (+) and 537~\AA\ (*) and \ion{He}{ii} $\lambda$\,304~\AA ($\Diamond$). Same model as for Fig.~\ref{profcrdprd}.}
  \label{linecore}
\end{figure}

The \ion{He}{i} 537 line exhibits the same type of variation as the 584 \AA\ line, with an interesting additional feature. When the prominence is at rest, the emergent intensity at line centre is actually less than the incident intensity. Although the main formation mechanism of the radiation still is the scattering of the incident radiation, the interlocking between the atomic states of \ion{He}{i} plays a key part in this case. When the upper state of the transition is excited through the incident radiation at that wavelength, it may come back to a lower energy state by other routes than the direct re-emission of a photon at 537 \AA. For instance, there might be a process involving photo-ionisation of the upper level of this transition followed by radiative cascade to the ground level.

\begin{figure}
  \resizebox{\hsize}{!}{\includegraphics{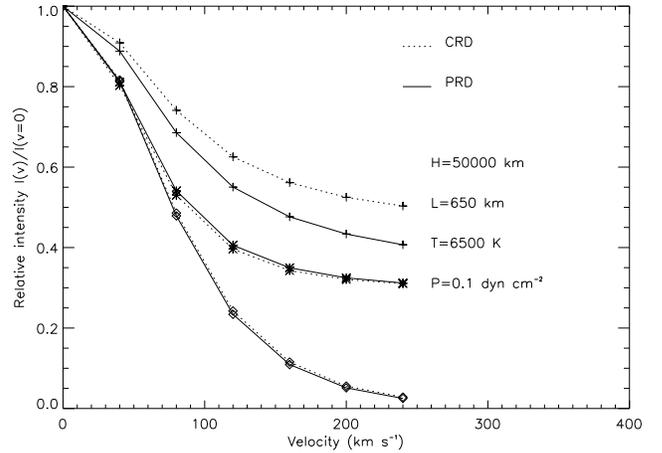}}
  \caption{Comparison of the relative integrated intensities {(integrated intensities normalised to the intensity at rest)} between CRD (dotted line) and PRD (solid line) calculations for \ion{He}{i} $\lambda\lambda$\,584 (+) and 537~\AA\ (*) and \ion{He}{ii} $\lambda$\,304~\AA ($\Diamond$). Same model as in {Fig.~\ref{profcrdprd}}.}
  \label{crdprdhe}
\end{figure}

The form of the frequency redistribution function {implies that most of} the differences between CRD and PRD {emergent line profiles} effectively arise {beyond 2--3} Doppler widths from the line centre \citep{hummer62,mihalas}. At 6500~K, {3 Doppler widths correspond} to approximately 30~m\AA, 28~m\AA, and 16~m\AA\ for the \ion{He}{i} 584, 537 and \ion{He}{ii} 304 lines respectively.
Figure~\ref{linecore} gives the relative contribution of the Doppler core for each emitted line (integrated intensity for wavelengths $\lambda_0-3\Delta\lambda_D \leq \lambda \leq \lambda_0+3\Delta\lambda_D$ normalised to the total integrated intensity of the line) {as a function of radial velocity}. In the CRD case (dotted line) {as in the PRD case (solid line)}, the most important contribution for the \ion{He}{i} 584 line intensity comes from the wings: 63\% {(54\%)} at zero velocity {in CRD (PRD)}, up to 82\% {(67\%)} at 240~km\,s$^{-1}$. {On the other hand,} the Doppler core has a predominant contribution for the \ion{He}{i} 537 and \ion{He}{ii} 304 lines. For these latter lines,  PRD effects in the line wings (where the intensity is already low) are therefore less noticeable, which explains the behaviour of {the line profiles (Fig.~\ref{profcrdprd}) and of} the relative intensities (Fig.~\ref{crdprdhe}). Thus the effect of PRD will be more important for the {\ion{He}{i}} line at 584 \AA. {For this line, the far wing contribution is decreased in PRD compared to CRD, but it remains predominant in the total emitted intensity.}

\begin{figure}
  \resizebox{\hsize}{!}{\includegraphics{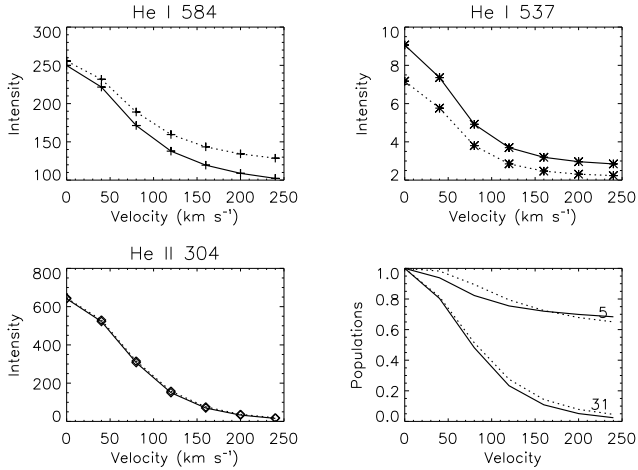}}
  \caption{{Emergent intensities in erg s$^{-1}$ cm$^{-2}$ sr$^{-1}$ as a function of radial velocity for \ion{He}{i} $\lambda\lambda$\,584 (upper left panel) and 537~\AA\ (upper right) and \ion{He}{ii} $\lambda$\,304~\AA (lower left). The lower right panel shows the variation of the relative populations (populations normalised to the population at zero velocity) with speed for levels {$^1$S\,1s2p of \ion{He}{i}} (5: upper level of the resonance transition at 584 \AA) and {$n=2$ of \ion{He}{ii}} (31: first excited state of \ion{He}{ii}, \textit{ie.} upper level of the resonance transition at 304 \AA). Dotted lines are for CRD computations and solid lines for PRD computations. Same model as in {Fig.~\ref{profcrdprd}}.}}
  \label{intpop}
\end{figure}

{Figure~\ref{crdprdhe}} shows the relative intensities (integrated intensities normalised to the intensity at zero velocity) as a function of the outward velocity of the structure for the same prominence models. Note that the effect of outward motion is a Doppler dimming effect on our three helium lines. This is expected as the incident lines are emission lines. 
Fig.~\ref{crdprdhe} shows that the effect of frequency redistribution {on the relative intensities} is especially important for the first resonance line of \ion{He}{i} at 584~\AA\ {and we have seen that this is mainly due to the wing contribution in the emitted intensity. The line }relative intensity in the PRD calculations is smaller than in the CRD case, and the difference between the two increases with velocity. For the {\ion{He}{i} 537~\AA\ and \ion{He}{ii} 304~\AA}\ lines, CRD and PRD lead to similar relative intensities at all velocities{, indicating that the variation with radial velocity of the balance between the different formation mechanisms of these lines is not affected by the details of the frequency redistribution}.

{However{, as already shown by the examination of the spectral line profiles in Fig.~\ref{profcrdprd}, }the absolute intensity of the \ion{He}{i} 537 line is different in PRD than in CRD. In figure{~\ref{intpop},} we show the variation of the absolute intensities of the three lines under study as a function of speed for both CRD and PRD cases, together with the variation of the mean populations of the upper levels of the resonance transitions at 584 \AA\ and 304 \AA. We do not show the variation of the population of the upper level of the resonance transition at 537 \AA\ as it is fairly similar to the variation of the upper level of the 584 \AA\ line. Whatever the type of the frequency redistribution, the effect of Doppler dimming on the resonance lines is a decrease of the populations of the upper states of these transitions. This is particularly striking in the case of the population of the first excited state of \ion{He}{ii}. The effect of the radial speed of the prominence material on the mean populations of the \ion{He}{i} singlet states involved in the resonance transitions is especially important between 50 and 150 km s$^{-1}$. At lower speeds, the incident radiation is still in resonance with the line absorption profile. At higher speeds, other mechanisms such as collisional excitations and atomic states interlocking prevent the fall-off of the population densities. }

{The consideration of the variation of the relative intensities with the radial velocity of the prominence allows us to better identify the parts played by the scattering of the incident radiation on one hand, and other processes such as thermal excitation on the other hand.}
The main conclusion is that it is important to compute the helium spectrum in PRD if we want to compare the calculations with observations {and infer plasma parameters in moving prominences. Still, CRD seems to be a good approximation for the \ion{He}{ii} line, despite the fact that the redistribution for this line is governed by quasi-coherent scattering. The explanation lies in the fact that the wing intensities are too low to notice the difference between CRD and PRD.}

\subsection{Sensitivity to temperature}

We now present PRD computations to show the {sensitivity on} the temperature {of} the {relative} integrated intensities and {absolute} line profiles for different velocities of the plasma. 
For this we compute {several} prominence models at two different temperatures (8000~K and 15\,000~K) for velocities between 0 and 400~km~s$^{-1}$, keeping other parameters fixed. These models are similar to the GHV models \citep{ghv} presented by \cite{gvg97b,gvg97a}.

\begin{figure}
  \resizebox{\hsize}{!}{\includegraphics{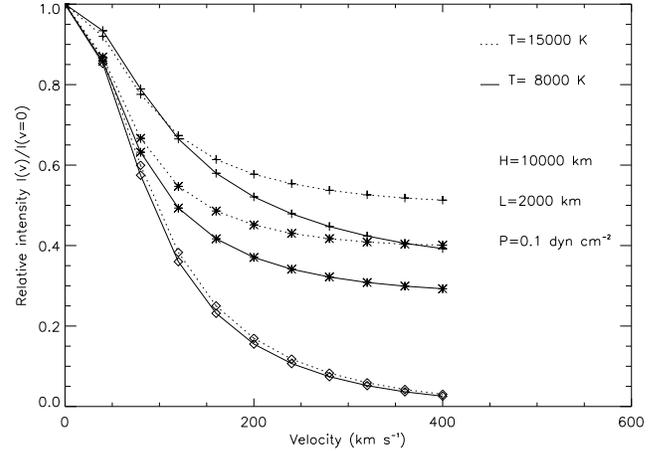}}
  \caption{Relative intensity as a function of velocity at 8000~K (solid line) and 15\,000~K (dotted line) for \ion{He}{i} $\lambda\lambda$\,584 (+) and 537~\AA\ (*) and \ion{He}{ii} $\lambda$\,304~\AA\ ($\Diamond$).}
  \label{tempint}
\end{figure}

\onlfig{9}{
\begin{figure*}
  \centering
  \includegraphics[width=15cm]{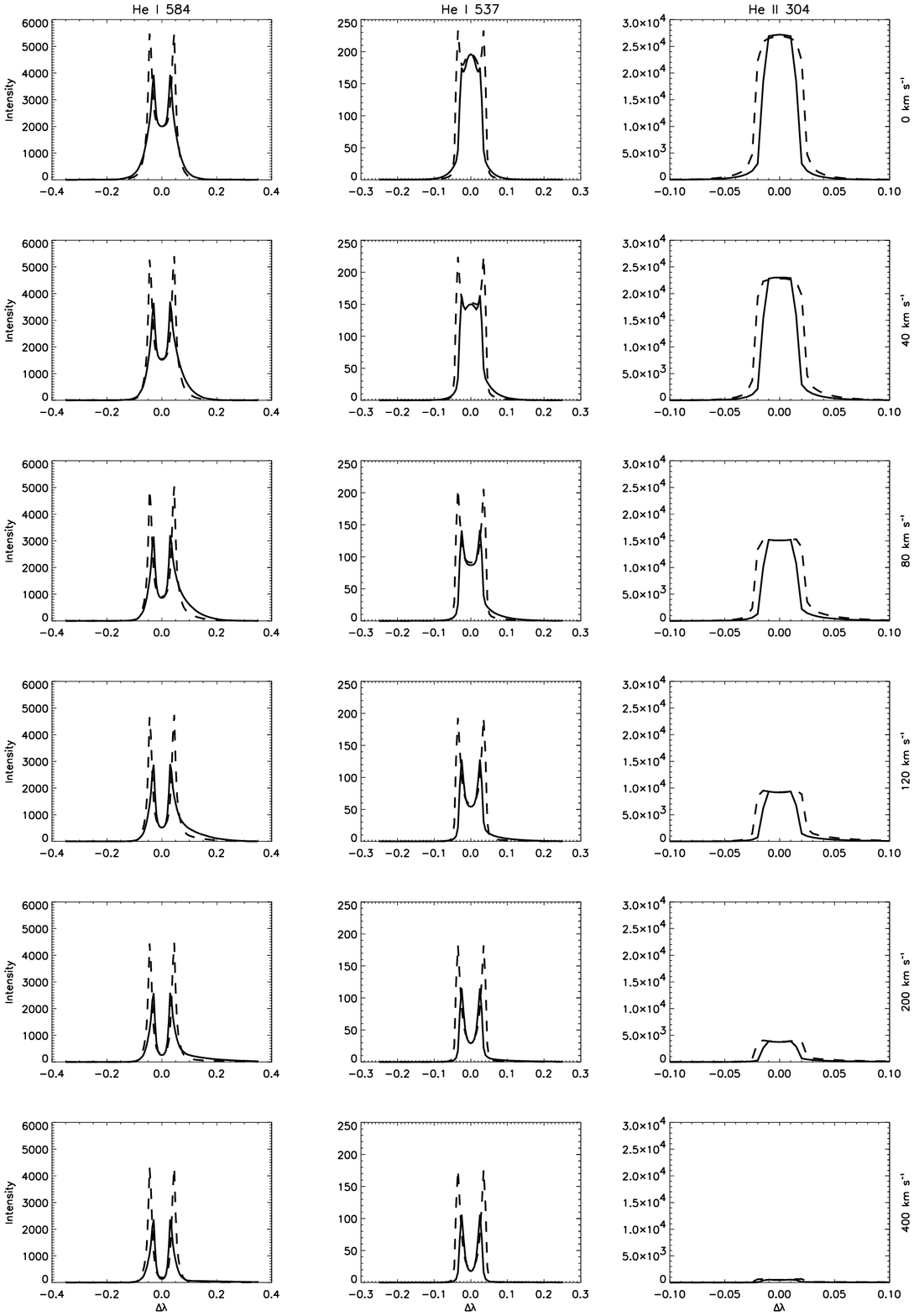}
  \caption{Line profiles for $T=8000$~K (solid line) and $T=15\,000$~K (dashed line), with $p=0.1$~dyn\,cm$^{-2}$, and $L=2000$~km, at different velocities: 0, 40, 80, 120, 200 and 400 km\,s$^{-1}$ from top to bottom. Abscissa is $\Delta\lambda$ in~\AA\ and vertical axis is specific intensity in erg\,s$^{-1}$\,cm$^{-2}$\,sr$^{-1}$\,\AA$^{-1}$.}
  \label{proft}
\end{figure*}
}

Figure~\ref{tempint} shows the relative intensities of the three helium lines as a function of the velocity of the structure. One can see that there is little difference between the two temperatures for the \ion{He}{ii} line. {In other words the integrated intensity of the \ion{He}{ii} line at 304 \AA\ is affected by velocity in a similar way at 8000~K and at 15\,000~K.} At these temperatures, the contribution of collisional excitation is negligible in this line \citepalias{lg01}. 

\begin{figure}
  \resizebox{\hsize}{!}{\includegraphics{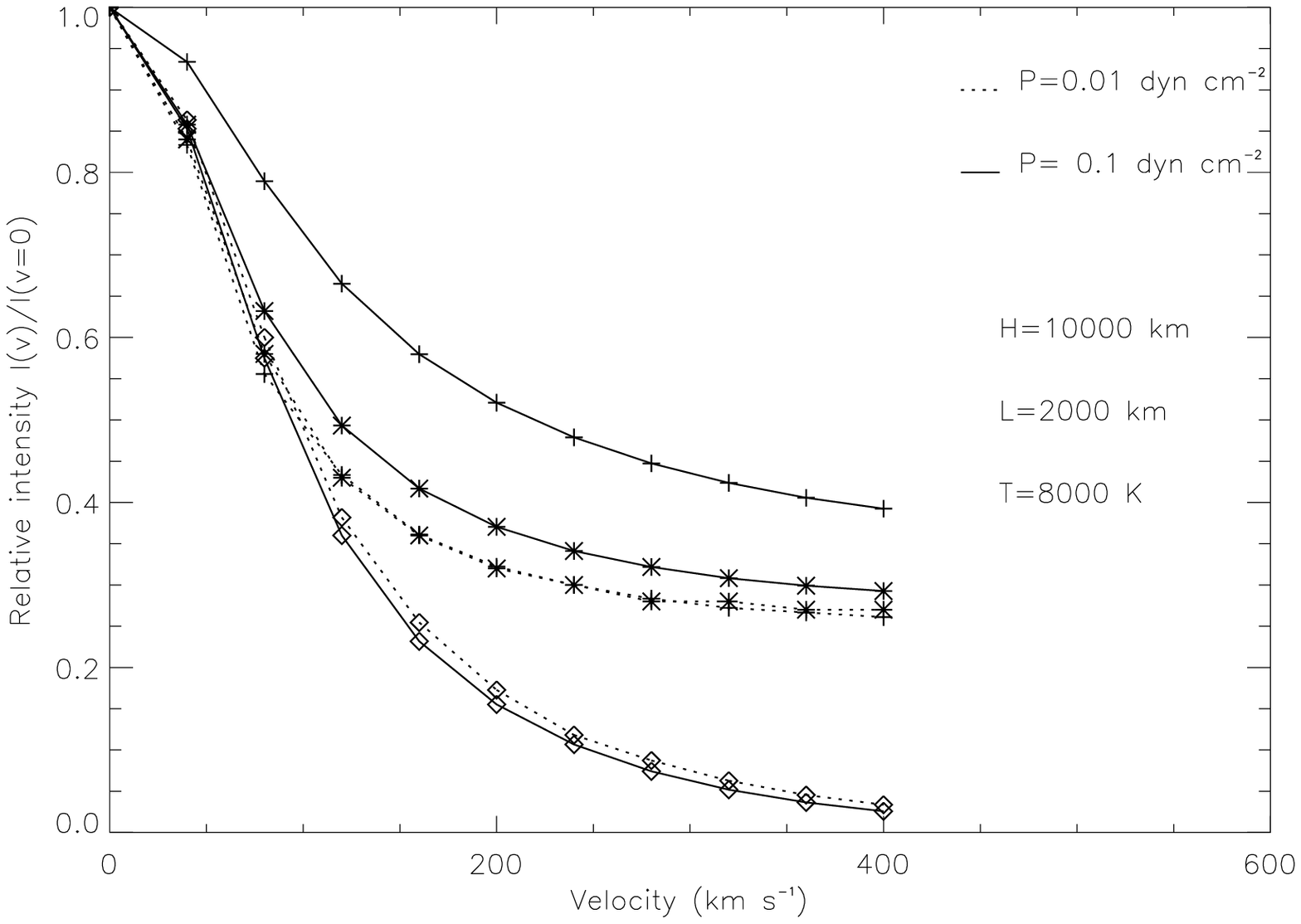}}
  \caption{Relative intensity as a function of velocity at 0.01~dyn~cm$^{-2}$ (dotted line) and 0.1~dyn~cm$^{-2}$ (solid line) for \ion{He}{i} $\lambda\lambda$\,584 (+) and 537~\AA\ (*) and \ion{He}{ii} $\lambda$\,304~\AA\ ($\Diamond$).}
  \label{presint}
\end{figure}

{The situation is different for the} \ion{He}{i} resonance lines, {since} the {relative} intensity profiles at 15\,000~K differ from those at 8000~K. The reason for that is twofold.
First, the increase in temperature broadens the absorption profile of the lines, thus effectively enhancing the amount of resonant scattering. Second, at high temperatures, the collisional excitation is not negligible for these lines. 
For each of the \ion{He}{i} line{s}, the difference{s} between the cold and hot prominence {in the line formation processes become} more evident as the velocity increases. 
This reflects the fact that the contribution of resonant scattering {in the formation of the lines} decreases with increasing velocity, hence the difference due to higher collisional rates is more visible at high speeds. 
At high velocities (greater than 150~km~s$^{-1}$), the relative intensity at 15\,000~K is greater than the relative intensity at 8000~K for all lines. Finally, at high temperatures, the {relative} intensities are less dependent on the velocity when the latter is high.
{We may need to stress that in Figs.~\ref{tempint}, \ref{presint}, and \ref{widthint}, the information on the effect of varying one particular parameter (temperature, pressure, or slab width) on the absolute values of the emergent integrated intensities is lost, as only relative intensities (normalised to the intensities at rest) are shown. However this information can be found in \citetalias{lg01}. }

The diagnostic of the plasma parameters is made possible by a careful study of the full line profiles. {Figure~\ref{proft} shows} the helium line profiles at different velocities for temperatures of 8000~K and 15\,000~K.
When the prominence is static, the emergent line profiles are symmetrical.
When the prominence is not static, the asymmetries of the profiles increase with speed, whatever the temperature is.
At a given speed, {the line integrated intensity is higher at 15\,000~K than at 8000~K, and the intensity at line centre is identical for both values of the temperature.
The latter arises from the fact that the plasma is strongly optically thick at these wavelengths at both temperatures, so that the line centre is saturated. Additional photons created by enhanced thermal processes, for instance, will only be able to escape the plasma after several scatterings that will shift their frequency away from the line centre.}

\subsubsection{\ion{He}{i} $\lambda$\,584~\AA}

At 8000~K the optical thickness at 584~\AA\ is higher than at 15\,000~K, because of the larger population of the ground level of \ion{He}{i} (lower level of the transition). While the core of the line is broader when the temperature increases, the intensity in the line wing decreases because there is less scattering.
For the low temperature and low speeds one notices an asymmetry in the line profile, with some intensity enhancement in the red part of the profile. For both temperatures the reversal at line centre is more pronounced with increasing velocity.
At high temperature the asymmetry in the line profile is less pronounced than at low temperature. The enhancement in the red wing compared to the blue one is best seen on a semi-logarithmic scale. The lower amount of thermal emission at 8000~K makes the signature of the velocity effect more visible than at 15\,000~K.
For all non-zero velocities, the incident profile, which is a single-peak emission line (see Fig.~\ref{incrad}), is quasi-coherently scattered in the red wing of the line.

\subsubsection{\ion{He}{i} $\lambda$\,537~\AA}

As for the 584 line, the intensity in the wings of the 537 \AA\ line decreases when the temperature is higher, because the opacity in the line is lower at 15\,000~K than at 8000~K. However the line core is broader at high temperature than at low temperature.
At 8000~K we see a central peak with two small peaks in the wings (due to high opacity) for the static model. When velocities are present, the central peak disappears and instead we find a reversal at line centre, which becomes more pronounced at higher velocities. 
The situation is the same at higher temperatures except that there is a self-reversal in the line core already for the static model.
The same remarks as for the 584~\AA\ line about profile asymmetries apply to the 537~\AA\ line: at a given temperature, the intensity in the red wing of the profile increases with velocity. At a given speed, the profile is more symmetrical at high temperatures than at low temperatures.

\subsubsection{\ion{He}{ii} $\lambda$\,304~\AA}

Helium ionisation is higher at 15\,000~K than at 8000~K and therefore the plasma is more optically thick for this line (as well as at the head of the \ion{He}{ii} continuum). Consequently the scattering of radiation in the line wings is enhanced when the temperature is higher, and the line core is broader.
The intensity of the \ion{He}{ii} 304 line is mainly dependent on the incident radiation, under the physical conditions considered in this study.

\subsection{Sensitivity to pressure}

We now look at similar prominence models ($T=8000$~K, $L=2000$~km) at two different pressures: $p=0.1$~dyn\,cm$^{-2}$ and $p=0.01$~dyn\,cm$^{-2}$. The relative intensities as a function of velocity are shown in Fig.~\ref{presint}, while Fig.~\ref{profp} presents the corresponding line profiles for velocities ranging between 0 and 400~km\,s$^{-1}$.
The asymmetries in the line profiles increase with velocity, but are similar at both pressures. 

\onlfig{11}{
\begin{figure*}
  \centering
  \includegraphics[width=15cm]{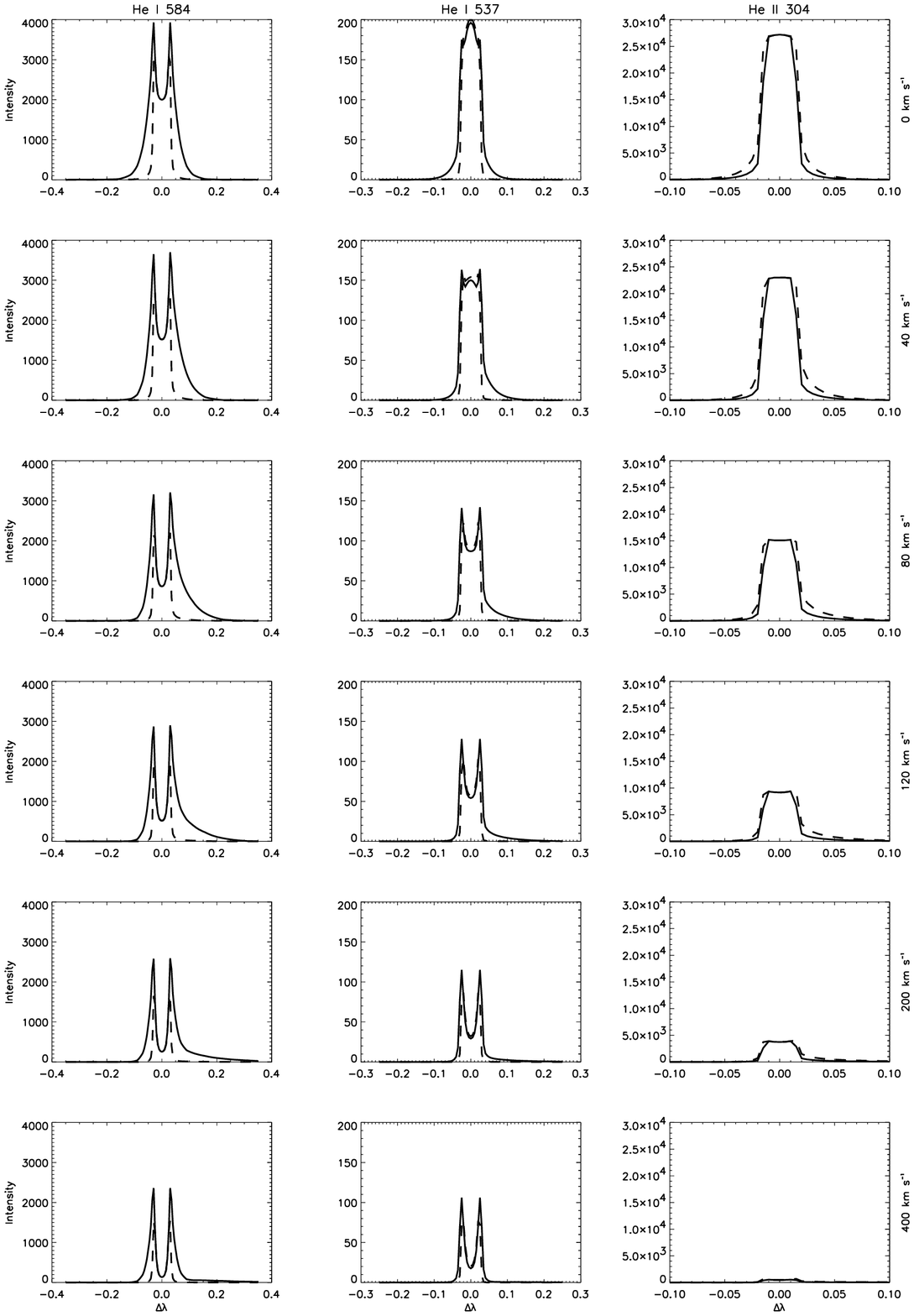}
  \caption{Line profiles for $p=0.1$~dyn\,cm$^{-2}$ (solid line) and $p=0.01$~dyn\,cm$^{-2}$ (dashed line), with $T=8000$~K and $L=2000$~km, at different velocities (same as in Fig.~\ref{proft}).}
  \label{profp}
\end{figure*}
}

\begin{figure}
  \resizebox{\hsize}{!}{\includegraphics{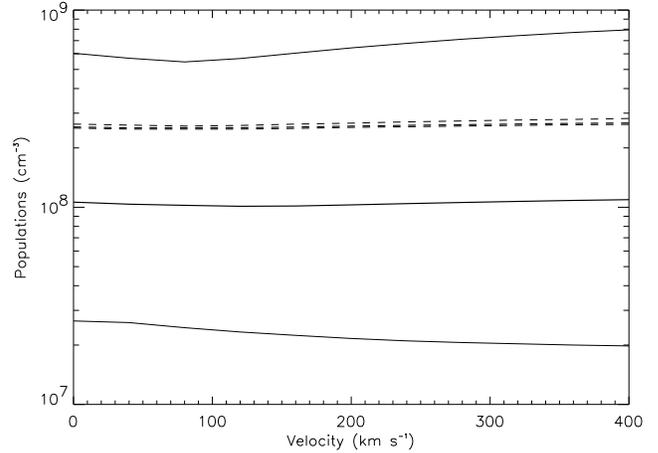}}
  \caption{{Population density of the ground level of \ion{He}{ii} as a function of velocity. The thickest lines refer to the mean population, the medium-thick line shows the level population at the surface of the slab, and the thinnest line corresponds to the population at slab centre. Solid lines: reference model with $P=0.1$~dyn cm$^{-2}$; dashed lines: low-pressure model with $P=0.01$~dyn cm$^{-2}$.} }
  \label{popint}
\end{figure}

\begin{figure}
  \resizebox{\hsize}{!}{\includegraphics{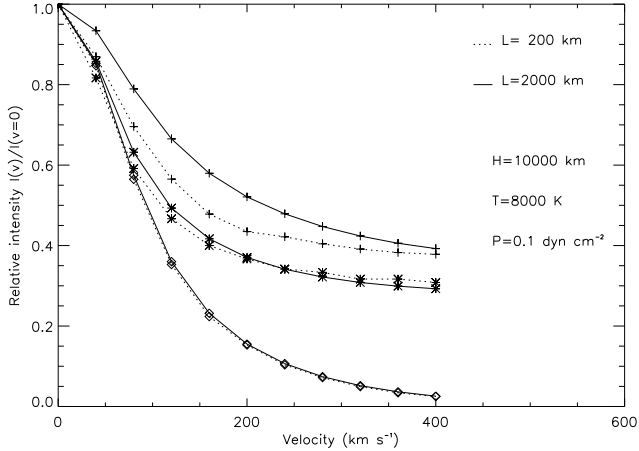}}
  \caption{Relative intensity as a function of velocity for a slab width of 200~km (dotted line) and 2000~km (solid line) for \ion{He}{i} $\lambda\lambda$\,584 (+) and 537~\AA\ (*) and \ion{He}{ii} $\lambda$\,304~\AA\ ($\Diamond$).}
  \label{widthint}
\end{figure}

\onlfig{14}{
\begin{figure*}
  \centering
  \includegraphics[width=15cm]{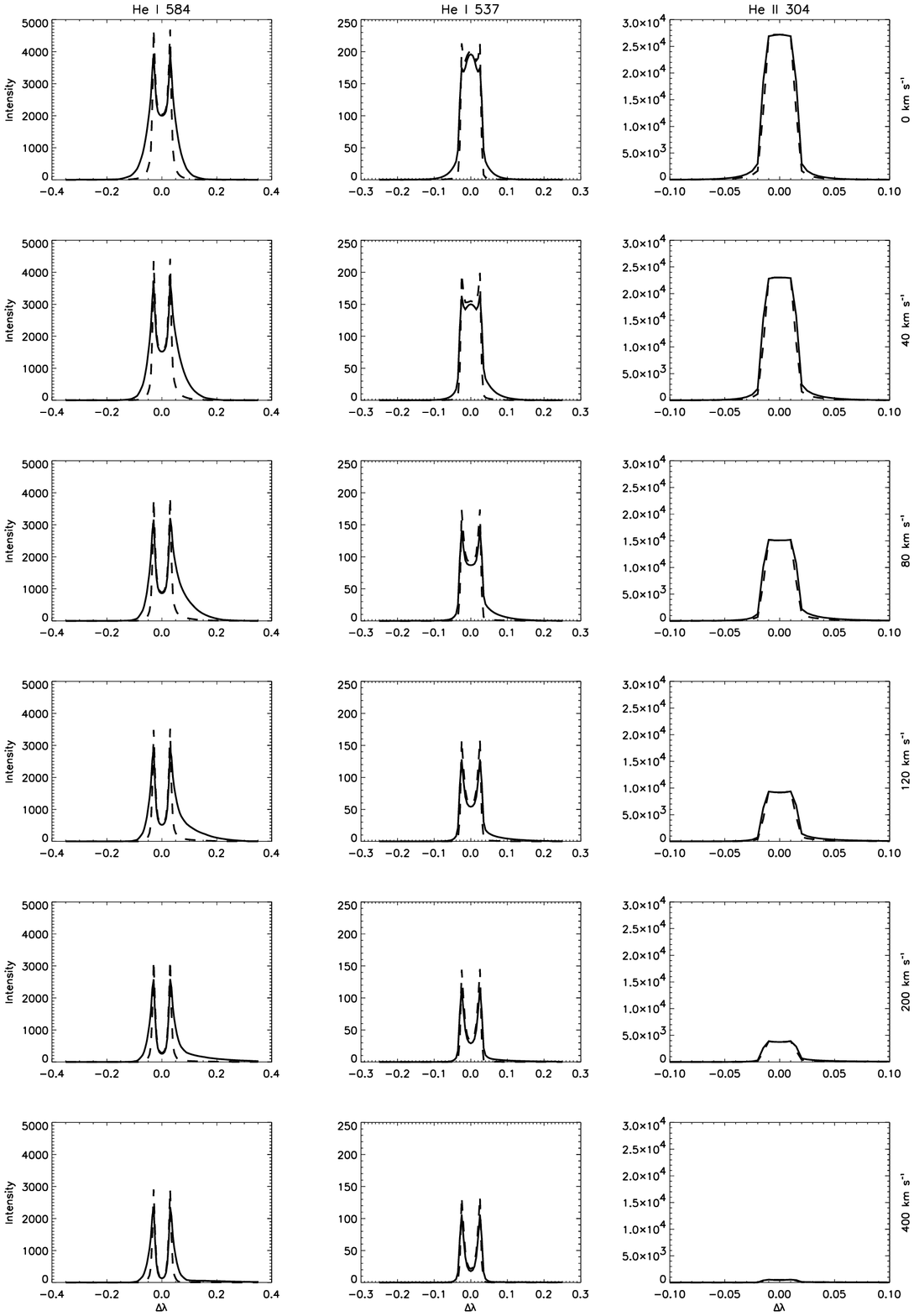}
  \caption{Line profiles for $L=2000$~km (solid line) and $L=200$~km (dashed line), with $T=8000$~K and $p=0.1$~dyn\,cm$^{-2}$, at different velocities (same as in Fig.~\ref{proft}).}
  \label{profw}
\end{figure*}
}

Fig.~\ref{presint} shows that the relative intensities of the \ion{He}{i} resonance lines are greater at higher pressure, while the opposite is true for the \ion{He}{ii} line.
An interesting feature is that at low pressure, the relative intensities are the same for the 584 and 537~\AA\ lines. This reveals very similar mechanisms of formation for the two lines at low pressures, dominated by radiative processes because of lower electron densities and of the plasma being optically thin at the head of the \ion{He}{i} resonance continuum (504~\AA). The increase of the radial velocity affects both line intensities in the same way.
When the pressure is higher, {the contribution of collisions in the formation of the lines becomes more important,} and the relative intensities of the two resonance lines become different.
The line profiles of the \ion{He}{i} lines are narrower at low pressure than at high pressure {(Fig.~\ref{profp})}, a consequence of lower collisional broadening and lower optical thickness. For all values of the velocity, the increase of pressure leads to enhanced scattering in the wings of these lines, while the line core is saturated.

The pressure has only a small effect on the relative intensity of the \ion{He}{ii} 304~\AA\ line, and indeed on the line profile. We explain it as follows: as the pressure is increased, the helium density is enhanced. In particular the population of the ground state of \ion{He}{ii} is larger at the edge of the prominence slab, while it is smaller in the main prominence central part {(Fig.~\ref{popint})}. Therefore the decrease of opacity in the 304 \AA\ line is almost compensated by an increase in the \ion{He}{ii} ground state population at the slab surface where the scattering of the incident radiation takes place. Consequently the intensity in the line wings is not drastically affected by the pressure variation. The intensity in the line core is not strongly modified by the change in pressure at this temperature: the profiles are slightly broader at low pressure. For both values of the pressure, this line is mainly formed by the resonant scattering of the incident radiation, leading to a strong sensitivity to the velocity of the plasma.
Figure~\ref{popint} also {shows} that at low pressures, the population density of the \ion{He}{ii} ground level is the same everywhere in the prominence slab, and is not affected by velocity.

\subsection{Sensitivity to slab width}

Finally, we study the effect of varying the slab width to represent different types of prominence threads. Keeping other parameters fixed to $T=8000$~K and $p=0.1$~dyn\,cm$^{-2}$, we compute the emergent line profiles for two slab widths: $L=200$~km and $L=2000$~km. The outward velocity varies again from 0 to 400~km\,s$^{-1}$. The relative intensities are shown in Fig.~\ref{widthint} and the line profiles on Fig.~\ref{profw}.

Fig.~\ref{widthint} shows that the relative intensities of the three EUV lines studied here respond to the variation of the width of the prominence slab in different ways. 
The relative intensity profile of the \ion{He}{ii} 304 line is virtually identical in the thin and the thick cases{, and indeed, the \ion{He}{ii} 304 line profiles for the thin and thick  prominences are very similar (Fig.~\ref{profw})}. In fact, the contribution of the radiative component in the source function of the line is not affected by the change in slab width. Therefore the velocity signature is identical in both cases.

The \ion{He}{i} lines relative intensities show more sensitivity to the thickness of the prominence slab. The relative intensity of the 584 line is particularly sensitive to the slab width at low speeds. At high velocity the relative intensities for the thin and thick cases are almost identical. Also, it can be seen that the {relative intensity of the} 584~\AA\ is not strongly dependent on the plasma velocity above 200~km\,s$^{-1}$ when the slab width is small.
The properties of the relative integrated intensity of the \ion{He}{i} 584 line with the slab width, and the differences with the 537 \AA\ line, are explained by the fact that for the 584~\AA\ line, the wings are the dominant contributor to the emitted intensity. The wings are formed more deeply in the prominence slab, while the line core is formed closer to the surface.
The asymmetries in the line profiles due to the Doppler dimming effects are similar for a slab 200~km thick and for a 2000~km thick slab. In both cases, the increase of the velocity increases the reversal at line centre and the asymmetry.

\section{Conclusions}\label{discuss}

{In this paper we have studied the effect of radial motions in prominences on the helium resonance lines \ion{He}{i} 584~\AA, \ion{He}{i} 537~\AA, and \ion{He}{ii} 304~\AA.}

The consideration of partial redistribution in frequency (PRD) is necessary to compute the profiles of the helium resonance lines emitted by moving material in prominences. This moving material could correspond to upwards/downwards mass flows in active prominences or ejected plasma in erupting prominences.

{The study of the variation of the relative intensities of the helium resonance lines with the radial velocity of the plasma indicates that th}e Doppler dimming effects are essentially present when the relative contribution of thermal emission compared to the scattering of incident radiation in the lines studied is low. 
At velocities larger than 150~km\,s$^{-1}$, most prominences show a strong sensitivity to the motion of the plasma in the \ion{He}{ii} 304~\AA\ line, and to a lesser extent, in the \ion{He}{i} 584~\AA\ and 537~\AA\ lines. At lower velocities, all \ion{He}{i} and \ion{He}{ii} EUV resonance lines are sensitive to Doppler dimming.

It has been shown in this paper that most of the information on the radial velocity can be obtained through the variation of the {relative} integrated intensity, \textit{ie.} by measuring the amount of Doppler dimming on the EUV resonance lines of helium. 
{Although this measurement is difficult to perform (mainly because one has to be observing at the right place and at the right time), the current space-based instruments such as EIT, SUMER, and CDS on SOHO, as well as TRACE and Hinode, are able to obtain a temporal series of intensity measurements covering an entire prominence eruption, starting before the radial motions take place. In this case it is possible to measure the dimming in intensity, and to compare it with our predicted relative intensity curves.}

{In a future study, a} more comprehensive diagnostic of the prominence plasma will be done through a comparison between the observed and our computed line profiles. Detectors with a spectral pixel size around 0.05 \AA\ or less (e.g. the SUMER instrument on SOHO) are ideal as they allow line asymmetries to be observed. A lower spectral resolution still allows good comparison with theoretical line profiles, however it makes the task of finding a unique model that fits the observed optically thick profiles more difficult. The line intensity of \ion{He}{ii} 304 may be used without knowing {the line} profile for measuring the radial speed of the erupting prominence.
In combination with the study of the apparent motion of the prominence material brought by {imagers on SOHO, Hinode, and STEREO,} the full velocity vector will be inferred.

   We have developed a powerful diagnostic tool that will significantly improve our understanding of the thermodynamical parameters of active and eruptive prominences. Using this set of three helium lines (\ion{He}{i} $\lambda\lambda$\,584 and 537~\AA\ and \ion{He}{ii} $\lambda$\,304~\AA) allows to constrain the models in order to reproduce the observations. Adding more lines such as hydrogen lines will yield even better constraints. We will better understand the differences between \textit{typical} erupting prominences showing the same morphology in \ion{He}{ii} and in H$\alpha$ \citep{wfb86}, and \textit{abnormal} erupting prominences showing extensive emission in \ion{He}{ii} but nothing in H$\alpha$ \citep{wangetal98}. This will be the subject of a future work.

\begin{acknowledgements}
  {We are indebted to the anonymous referee who made several useful comments which helped to improve the clarity of this paper.}
NL acknowledges support from PPARC grant PPA/G/O/2003/00017.
\end{acknowledgements}


\end{document}